# Adapt/Exchange decisions depend on structural and surface features: Effects of solution costs and presentation format

Romy Müller

*Faculty of Psychology, Chair of Engineering Psychology and Applied Cognitive Research, TUD Dresden University of Technology, Dresden, Germany*

Corresponding author:

Romy Müller

Chair of Engineering Psychology and Applied Cognitive Research

TUD Dresden University of Technology

Helmholtzstraße 10, 01069 Dresden, Germany

Email: romy.mueller@tu-dresden.de

Phone: +49 351 46335330

Fax: +49 351 46337741

ORCID: 0000-0003-4750-7952

# Abstract

Problem solvers often need to choose between adapting a current solution and exchanging it for a new one. How do such decisions depend on structural and surface features of the task? The present study investigated the interplay between the costs of the two solutions (a structural feature) and the format in which this information was presented (a surface feature). In a computer-based modular plant scenario, participants chose between process parameter modifications (Adapt) and reconfigurations of the module setup (Exchange). Solution costs were presented either as graphs depicting parameter relations, separate numbers for each parameter, or integrated numbers for each solution. It was hypothesised that graphs induce satisficing (i.e., basing decisions only on Adapt), whereas the numeric formats foster a comparison of the solutions (i.e., basing decisions on the Adapt/Exchange ratio). The hypothesised effects were restricted to situations with medium Adapt costs. A second experiment replicated these findings while adjusting the scale of numeric formats. We conclude that Adapt/Exchange decisions are shaped by an interaction of structural and surface features.

# 1. Introduction

The balance between stability and flexibility is a cornerstone of adaptive action control (Goschke, 2013; Hommel, 2015). However, human decision makers can be remarkably inflexible. They tend to go with defaults (Jachimowicz et al., 2019) and try to keep up the status quo (Gal, 2006; Ritov & Baron, 1992; Samuelson & Zeckhauser, 1988). However, sometimes changes are inevitable, because a previously successful solution does not work anymore. In such situations, what kind of changes do people make? Do they merely adapt the details of the current solution, or exchange it for a completely different solution principle? In industrial contexts, the technical innovation of modularity introduces this issue into the daily work of operators, as modular plants enable a flexible reconfiguration of their physical setup (Nyhuis et al., 2008). Accordingly, operators can choose between local and global changes (Müller & Urbas, 2017): They can either modify process parameters like temperature or pressure within the narrow ranges of the current module (Adapt), or reconfigure the plant and use another module that enables production at more suitable parameter ranges, for instance because it has a bigger reactor (Exchange). How do people arrive at a decision in such situations?

## 1.1 Decision strategies in Adapt/Exchange scenarios

The cognitive processes underlying Adapt/Exchange decisions are poorly understood. One open question is under what conditions people engage in a thorough comparison of both solutions. Two previous studies suggested that decision makers often refrain from checking the alternative Exchange solution (Müller & Pohl, 2023) and only base their decisions on the costs of Adapt, while ignoring the cost ratio of Adapt and Exchange (Müller & Urbas, 2020). In other words, they seem to apply a satisficing strategy (Simon, 1956, 1990). Satisficing is defined as "using experience to construct an expectation of how good a solution we might reasonably achieve, and halting search as soon as a solution is reached that meets the expectation" (Simon, 1990, p.9). While previous research has often understood satisficing as a specific heuristic during sequential choice (e.g., Todd & Miller, 1999), we use the term in a broader sense: considering only a particular solution, as long as it is good enough. This understanding of satisficing aligns with its use in the framework of Naturalistic Decision Making (Klein, 2008): when experts solve emergency problems, they go with the first solution that comes to mind and passes the test of mentally simulating its outcome. Usually, no further alternatives are considered.

With regard to satisficing in previous Adapt/Exchange studies (Müller & Pohl, 2023; Müller & Urbas, 2020), it is an open question whether participants relied on this strategy because it was suitable given the structural features of the task. Alternatively, surface features might have nudged them into a heuristic type of processing. The former possibility rests on the assumption that simple heuristics like satisficing can lead to good outcomes when they are adjusted to the structure of the environment (Todd et al., 2012). One condition under which satisficing greatly supports decisions is when the options are incommensurable (Simon, 1990), either because their large number of attributes cannot be compared, or because their outcomes are uncertain. Both factors apply to Adapt/Exchange decisions, particularly in complex industrial settings. This is because the two solutions may not only differ in their explicitly quantified costs (e.g., magnitude of undesirable interference with the production process), but also in other, non-quantified costs (e.g., efforts and risks). Thus, the explicit attributes only represent a fraction of the actual attributes of each solution. These explicit and non-explicit costs represent a structural feature of Adapt/Exchange decisions, which might have induced satisficing in previous studies.

Alternatively, satisficing might have resulted from surface features of the task, particularly the format in which the Adapt and Exchange costs were presented. It might just have been too difficult to compare the two solution principles. If so, people should no longer satisfice when the presentation format facilitates a comparison of solutions.

Taken together, we currently do not know whether people satisficed because of the way the costs of the solutions were distributed (i.e., a structural task feature) or because of the way the information was presented (i.e., a surface feature). Therefore, the present study investigated how structural and surface features of Adapt/Exchange tasks interact in shaping decision strategies. Before introducing the experimental setting, we will provide a theoretical background on effects of presentation format in different decision contexts, and elaborate on the requirements of Adapt/Exchange decisions in a particular industrial scenario.

**1.2 How do presentation formats affect decision making?**

Effects of presentation format have been observed in several decision contexts, and two of them are particularly relevant to Adapt/Exchange decisions: intertemporal choice and multi-attribute decision making. In intertemporal choice or delay discounting, people choose between a sooner/smaller and a larger/later reward, and consistently discount rewards as a function of time. That is, rewards lose their subjective value when people have to wait longer to receive them (Berns et al., 2007; Frederick et al., 2002). The rate of such discounting varies with presentation format, and two types of format effects have been reported. One depends on the *units of rewards and delays*. In short, people are more willing to wait when the time interval is less transparent or when the amount of reward is more transparent. The time interval can be made less transparent by presenting delays as specific dates, rather than numbers of days (DeHart & Odum, 2015; LeBoeuf, 2006; Read et al., 2005). The rewards can be made more transparent by making them easier to estimate, for instance by using familiar units (DeHart et al., 2018). Both manipulations cause people to be more patient. The second type of presentation format effect relies on a *highlighting gains and losses*. In short, people are more willing to wait when the consequences of both options are emphasised and clearly differentiated. This can either be achieved by making explicit what is missed when not choosing the larger/later option (Magen et al., 2008; Radu et al., 2011), or by emphasising the costs of choosing the sooner/smaller option (Grace & McLean, 2005). Again, both manipulations cause people to be more patient. Taken together, depending on how the options are presented, attention can be guided to different features of the decision (e.g., length of the time interval, magnitude of the delayed reward of the larger/later option, unwanted future consequences of the sooner/smaller option). This differential attention allocation may change how people decide.

A second decision context in which presentation format influences choices is multi-attribute decision making. In this setting, people choose between two or more options that differ in their values on a number of shared attributes (e.g., four houses differ in their size, price, quality of the neighbourhood, connection to public transport). People usually do not compare all options on all attributes weighted by their respective relevance or validity, but use simpler heuristics (Bettman et al., 1993; Payne et al., 1988). Effects of presentation format come in two types. One depends on a *variation of the modalities and scales of attributes*. In short, the influence of attributes can be increased when they are made easier to comprehend or when they are expanded. Making attributes easier to comprehend can be achieved by supplementing numerical information with explicit quality evaluations (Peters et al., 2009), by using graphical formats instead of just numbers (Peters et al., 2009; Stone et al., 1997), and by expanding the axes of diagrams to make the respective attributes seem more important (Sun et al., 2010). Similar expansions can be applied to numbers, for instance by presenting prices per year instead of per month (Burson et al., 2009). All of these manipulations make people put more weight on the respective attributes, increasing their impact on decisions. The second type of presentation format effect depends on the *need for information search*. In short, when presentation formats make the information less accessible and the comparison of options more difficult, people rely on simple heuristics, instead of comparing the options on all attributes in a weighted additive manner. The accessibility of information is reduced when people have to go through the attributes sequentially (Glöckner & Betsch, 2008), or when a map-like presentation format ties each attribute to a particular location (Söllner et al., 2013). Both increase the need for information search and lead to simpler, more selective decision strategies. Taken together, how people use attributes depends on whether these attributes are comprehensible and accessible. People are more likely to base their decisions on information that they can easily find, understand, evaluate, and integrate.

Adapt/Exchange decisions are similar to both of the previously discussed decision contexts (for a detailed comparison see Müller & Urbas, 2020). For one, they require choices between a solution that is low in rewards and costs versus one that is high in rewards and costs, making them similar to intertemporal choice. For another, they require assessing and integrating a variety of different features of the two solutions, making them similar to multi-attribute decisions. This raises the possibility that some of the presentation format effects discussed above can also make people less inclined to satisfice and more inclined to compare the Adapt and Exchange solutions. Therefore, in the present study we aimed to make attribute values more or less easy to estimate and compare, and to facilitate or impair information search.

**1.3 Why do people satisfice when making Adapt/Exchange decisions in modular plants?**

Consider the following scenario. While producing an expensive chemical, a problem can either be tackled by adapting process parameters (e.g., temperature, pressure) or by exchanging the current reactor module for a more suitable, bigger one. Operators are informed that, if they choose Adapt, they need to substantially increase temperature in order to still reach the production goals, as the current reactor actually is too small. Such process interventions put the chemical process at risk and thus you generally want to keep them to a minimum. Instead, much less process intervention is required when choosing Exchange, because the bigger reactor of the new module makes it possible to substitute temperature changes for volume changes, which have no negative side effects. Therefore, Exchange seems preferable according to the available information. However, at the same time it has costs that are not made explicit: The module exchange requires physical effort, takes time, and can pose its own (partly unknown) risks for the process and plant. These non-explicit costs of Exchange might outweigh the costs of Adapt, especially when the latter only requires minor adjustments of process parameters. Thus, why would operators invest efforts in a module exchange, when the production goals can safely be achieved by merely adapting parameters? In some situations, it simply might not matter how much better Exchange performs on the explicit attributes (e.g., required process intervention), namely when the solution achievable via Adapt is already known to be good enough.

This could explain why in a previous study (Müller & Urbas, 2020), no evidence for a thorough comparison of the two solutions was found. Participants went through sequences of Adapt/Exchange decisions with gradually changing cost ratios (i.e., process intervention required for Adapt vs. Exchange). That is, Adapt either became successively better or worse than Exchange. It was also varied whether the gradual cost ratio changes were accompanied by gradual increases in the absolute Adapt costs, or whether these absolute costs alternated between trials. Participants' choices did not depend on the gradual increase in cost ratio, but only on the absolute costs of Adapt. Moreover, participants were faster when choosing Adapt than Exchange. These results suggests that Exchange choices involved an additional mental operation (i.e., evaluating the Exchange solution) and that participants skipped this additional effort when choosing Adapt. Apparently, they simply chose Adapt whenever it was good enough. That is, they adopted a satisficing strategy.

Satisficing might have been a suitable strategy, due to the structural features of the Adapt/Exchange decisions. However, it needs to be noted that participants had to base their decisions on a rather complicated graph visualisation, which depicted the costs of the two solutions as relations between process parameters and production goals. This presentation format arguably made the comparison of the solutions rather difficult. As high task difficulty can induce satisficing (Krosnick, 1991), it is unclear whether the previous findings were a consequence of excessive cognitive demands, while another presentation format might have led participants to engage in a thorough comparison of the solutions. This was investigated in the present study.

### 1.4 Present study

The present study compared three formats for presenting the costs of Adapt and Exchange, operationalised as the process intervention that was needed to achieve production goals. The first presentation format used the *graphs* from a previous study (Müller & Urbas, 2020): curves representing the thresholds for two production goals, and the distance in parameter space that needed to be covered to achieve these goals. These graphs provided information in a spatial format and required considerable effort of information search and integration to compare the costs between the two solutions. In previous work, similar requirements impaired information integration (Söllner et al., 2013). Second, *separate numbers* presented the costs of Adapt and Exchange in a numerical format, and separately for each process parameter. Accordingly, comparing the solutions made it necessary to calculate the total cost for each solution by adding up the component costs. A third presentation format, *integrated numbers*, presented the overall cost of each solution as a single number, and thus only required a comparison of two numbers. The aim of the present study was to investigate how these three presentation formats affect participants' decisions, leading them to either satisfice, or to compare the Adapt and Exchange solutions by computing their cost ratio. Alternatively, if a satisficing strategy is generally adopted when making Adapt/Exchange decisions, simply because it makes sense in this context, presentation format should have no effect. To investigate how the influence of presentation format as a surface feature depends on structural task features, two such features were varied: the absolute costs of Adapt, and the cost ratio of Adapt and Exchange. If participants satisfice, only the absolute costs of Adapt should matter, but not the Adapt/Exchange ratio. This was investigated in two experiments using different operationalisations of process intervention costs.

## 2. Experiment 1

### 2.1 Introduction

*2.1.1 Experimental setting*

In a computer-based experiment, participants were responsible for enhancing product quality in a chemical process. Product quality depended on two factors: conversion and foam (see Figure 1A). While conversion denotes how much of the chemical has reacted, and thus high conversion is an outcome to strive for, foam can occur as a side effect of chemical reactions and can destroy the product. Therefore, enhancing product quality made it necessary for participants to increase conversion, while avoiding foam. Moreover, they had to intervene with the process as little as possible, as any process intervention is risky. To reach the production goals, participants could adjust three process parameters: temperature, mixing speed, and volume. Volume and temperature could be used to increase conversion. While volume had no negative side effects, temperature also increased foaming. Therefore, temperature increases usually had to be compensated by reductions of mixing speed to avoid foaming. Goal conflicts arose from the fact that the required conversion could only be achieved by increasing volume and/or temperature, while temperature also increased foaming and thus was a non-desirable process intervention. While foam could be avoided via mixing speed, this compensation additionally increased the non-desirable process intervention (see Figure 1B).

**Figure 1**

Causal diagrams representing the constraints in the modular plant scenario. (A) Relations between process parameters and outcomes. (B) Goal conflicts in the selection of parameter settings.

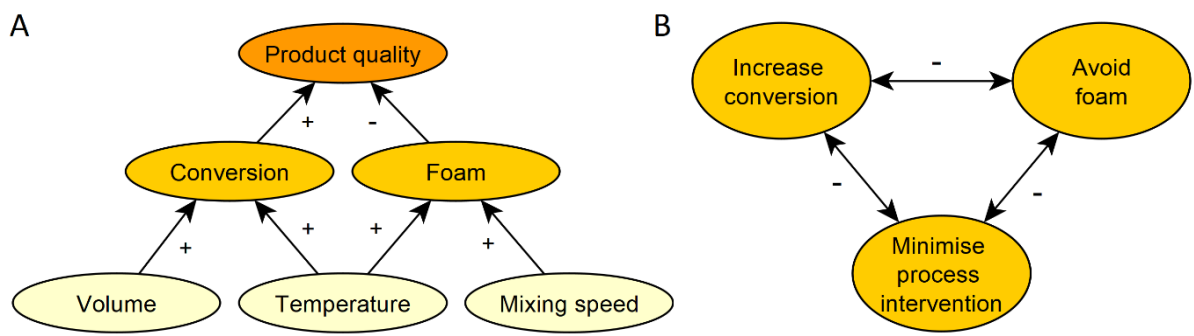

Participants could choose between adapting parameters in the currently used, small module, or exchanging this module for a new one with a bigger reactor. This module choice determined how process parameters could be set, because the two modules had different but overlapping operating ranges. These ranges were much narrower in the current module, and this determined how production goals could be achieved by mere parameter adaptations: the required conversion could not be reached by increasing volume, because the small reactor did not allow for this. Therefore, when choosing Adapt, participants had to use temperature and compensate its negative effects via mixing speed, which required a large process intervention. Conversely, when choosing Exchange, participants could increase volume to reach the required conversion, thus greatly reducing the changes in other parameters. However, Exchange came with an additional, non-explicit cost: the physical effort required to manually exchange the module. This cost was not included in the presentation made available to participants. Still, it was anticipated to shift participants' preferences towards choosing Adapt, although Adapt was inferior to Exchange in the explicit costs throughout the experiment. Thus, participants were confronted with a goal conflict: On the one hand, they had to minimise the explicit costs (i.e., process intervention) but on the other hand, they also had to keep the non-explicit costs (i.e., physical effort) at an acceptable level. There was no normatively correct solution, as solution quality depended on how participants weighted the explicit and non-explicit costs.

In the present study (and in contrast to Müller & Urbas, 2020), participants did not have to generate the parameter settings by themselves, but each trial provided one Adapt and one Exchange solution, which participants could choose from. Both solutions led to a successful outcome, but they differed in the costs of achieving these outcomes (i.e., larger process intervention for Adapt, smaller process intervention plus physical effort for Exchange).

*2.1.2 Presentation formats*

Three presentation formats were compared: graphs, separate numbers, and integrated numbers. Each provided information about the explicit costs of changing process parameters, but no information about the non-explicit costs of physically exchanging the module or the associated risks.

*Graphs* represented the relevant relations in the chemical process (see Figure 3A). The left curve shows how the current conversion threshold depends on temperature and volume. The right curve shows how the threshold for foam risk depends on temperature and mixing speed. All values above the left curve and below the right curve are acceptable. The graph visualisations also presented the position of the current parameter settings and their new positions given that Adapt or Exchange was chosen. In this way, participants could see how much process intervention they would need for each solution (i.e., to move from the current position to the new one). In the present study, all presented graphs provided valid solutions (i.e., all Adapt and Exchange solutions were located above the conversion curve and below the foam curve, respectively). Thus, graphs provided information about process intervention costs in spatial form. In that sense, they were similar to the complex map format used by Söllner et al. (2013): several information elements were spatially distributed, and participants had to search and integrate them in order to compare the solutions.

The second presentation format provided *separate numbers*: numeric values representing the costs of Adapt and Exchange, split up according to the parameters that had to be changed (see Figure 3C). Thus, for each solution a value was provided for temperature, one for mixing speed, and one for volume. Separate numbers corresponded to the graph visualisations in that they indicated the grid steps required for each parameter to move from the current position to the new one (see arrows in Figure 3A). Accordingly, comparing the costs of Adapt and Exchange required participants to calculate the sum of parameter changes for each solution.

The third presentation format provided *integrated numbers*: the total costs of each solution (i.e., number of required parameter steps), summed over the two relevant parameters of temperature and mixing speed (see Figure 3D). This presentation format made it easiest to compare Adapt and Exchange, as participants only had to compute the ratio of two numbers. In that sense, it was comparable to previous presentation formats that supported option evaluation, for instance by presenting delays as days instead of dates (DeHart & Odum, 2015; LeBoeuf, 2006; Read et al., 2005) or presenting rewards in clear instead of fuzzy units (DeHart et al., 2018).

*2.1.3 Adapt costs and Adapt/Exchange ratio*

The experimental trials differed with regard to the required Adapt costs (i.e., low, medium, high). At each level of Adapt costs, the respective Exchange costs were set to produce three Adapt/Exchange ratios (i.e., 6:1, 6:2, 6:3). The presumed consequences of different decision strategies are depicted in Figure 2. If participants satisfice, their percentage of Exchange choices should only depend on Adapt costs, but not on the Adapt/Exchange ratio (see Figure 2A): they should choose Adapt as long as it is good enough, regardless of how much better Exchange might be. Conversely, if they compare the two solutions, their choices should also depend on the Adapt/Exchange ratio (see Figure 2B), with more Exchange choices at higher Adapt/Exchange ratios (i.e., more for 6:1 than 6:3). Finally, participants' strategies of satisfying versus comparing solutions might depend on the costs of Adapt. In this case, their choices should be independent of the Adapt/Exchange ratio when Adapt costs are low, but become more dependent on the Adapt/Exchange ratio as Adapt costs increase (see Figure 2C).

**Figure 2**

Presumed effects of satisficing versus comparing solutions on the percentage of Exchange choices, depending on Adapt costs and Adapt/Exchange ratio. (A) Satisficing while completely ignoring the Adapt/Exchange ratio, (B) always considering the Adapt/Exchange ratio, and (C) increasingly considering the Adapt/Exchange ratio as Adapt costs increase.

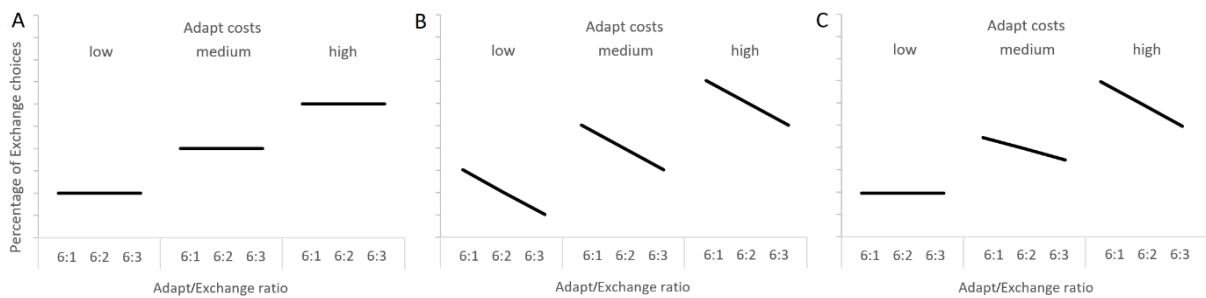

*2.1.4 Hypotheses*

With graphs, we expected that participants would satisfice (i.e., choose Adapt when it is good enough, regardless of the quality of Exchange). Conversely, with the numeric presentation formats, we expected them to compare the two solutions, and thus also take the Adapt/Exchange ratio into account. Accordingly, the critical comparison was whether the percentage of Exchange choices would differ between the highest and lowest Adapt/Exchange ratio. Such effects of Adapt/Exchange ratio were expected to be weak or absent with graphs, but strong with numbers.

We also hypothesised that participants would choose Exchange more often when Adapt costs increase. More important than this main effect, we expected a triple interaction between presentation format, Adapt/Exchange ratio, and Adapt costs: whether presentation formats are differentially sensitive to the Adapt/Exchange ratio (as an indicator of solution comparison) might depend on the costs of Adapt. With graphs, we expected participants not to compare solutions when Adapt is good enough, but to compare solutions when Adapt is problematic. Thus, Adapt/Exchange ratio effects should be absent at low and medium Adapt costs, but present at high Adapt costs. With the numeric presentation formats, we expected participants to generally compare solutions as a default strategy. Thus, Adapt/Exchange ratio effects should be present at all levels of Adapt costs. This also means that we only expected presentation format to affect strategy choice when Adapt costs are sufficiently small. That is, numbers but not graphs should show effects of Adapt/Exchange ratio at low and medium Adapt costs.

We had no specific hypotheses for differences between the two numeric formats. Both separate and integrated numbers were expected to increase solution comparison (and thus Adapt/Exchange ratio effects) compared to graphs. However, we were not sure whether the higher integration difficulty of separate numbers would impair solution comparison, relative to integrated numbers. If so, this should result in lower Adapt/Exchange ratio effects for separate numbers than integrated numbers.

## 2.2 Methods

*2.2.1 Participants*

Twenty-six members of the TUD Dresden University of Technology participant pool (ORSEE, Greiner, 2015) took part in the study in exchange for course credit or 7€ per hour. One participant was excluded from the data analysis as she did not pass the knowledge test administered after the instruction video (see below). Thus, the final sample consisted of 25 participants (15 female, 10 male) with an age range of 19 to 68 years (*M* = 28.8, *SD* = 10.6). Participants provided written informed consent, and all procedures followed the principles of the Declaration of Helsinki.

*2.2.2 Apparatus and stimuli*

*Instruction video.* Before starting the experiment, participants watched an instruction video that was taken from a previous study (Müller & Urbas, 2020). The video was based on a Microsoft PowerPoint presentation, lasted 17 min, and used several instructional techniques to facilitate learning (e.g., advance organisers, animations, summary slides, and test questions). It consisted of three parts:

- Explanation of the chemical process with a focus on the causal relations between process parameters (i.e., volume, temperature, mixing speed) and outcomes (i.e., conversion, foam)

- Introduction to modular plants, characteristics of the small and big module with regard to the process parameters, and positive/negative effects of Adapt and Exchange

- Instruction concerning the materials and decisions in the experiment, as well as the following rules of thumb: (1) Parameter changes are risky for the product and thus you should change as few parameters as possible, and change each parameter as little as possible. (2) Volume does not harm the process. (3) Temperature is the parameter with the strongest positive and negative effects. (4) Usually, there is more than one correct solution.

The last part provided an instruction for the experiment. It explained the stimuli in detail, and used an animated example as a step-by-step demonstration of how the new parameter values in graphs were determined. Specifically, participants were told that conversion was reached when the position of the x was above the left curve, which meant that temperature had to be increased from value $T_1$ to value $T_2$. It was also stated that as a consequence of this increase, the x in the right picture moved to $T_2$ as well, and therefore its new position was above the foam curve. Therefore, it was necessary to reduce mixing speed from value $S_1$ to $S_2$. The instruction did not refer to steps of process intervention, but only indicated the initial and final parameter values. It also did not suggest any strategies for choosing between Adapt or Exchange. Instead, it used the same stimulus to first demonstrate the procedure given that Adapt was chosen, and then a second time given that Exchange was chosen.

*Experiment.* The experiment took place in a quiet lab room, where up to three participants worked in parallel during each session, using one of three laptops (13, 14 and 15.6", respectively) and a standard computer mouse as an input device. The experiment was programmed with the Experiment Builder (SR Research, Ontario, Canada). Example stimuli are presented in Figure 3. All stimuli were displayed at a resolution of 1920 × 1080 pixels. They presented pictures, interaction elements, and text in white font on a black background, and consistent colour coding was used for Adapt/small module (i.e., green) and Exchange/big module (i.e., purple). All text was presented in German. Each stimulus included the following elements: (1) a textual request to achieve conversion and avoid foam, (2) the actual graph or number stimulus, (3) a legend explaining the contents of the graph or number stimulus, as well as (4) a green button to choose Adapt and a purple button to choose Exchange.

**Figure 3**

Stimulus material. (A) Relations between process parameters and outcomes for one trial. Green and purple background areas represent the parameter ranges of the small and big module, curves represent the thresholds for achieving conversion (left picture) and avoiding foam (right picture). An x marks the current parameter values, and the green and purple dots preview the new values for Adapt and Exchange, respectively. The Adapt/Exchange ratio is 6:2, which is generated as follows. For Adapt, six steps in the grid are needed to reach the green dot (see green arrows): four for temperature and two for mixing speed. For Exchange, two steps are needed to reach the purple dot (see purple arrows): two for temperature and zero for mixing speed. The final three parts of the figure show example screens for each presentation format: (B) graphs, (C) separate numbers, (D) integrated numbers.

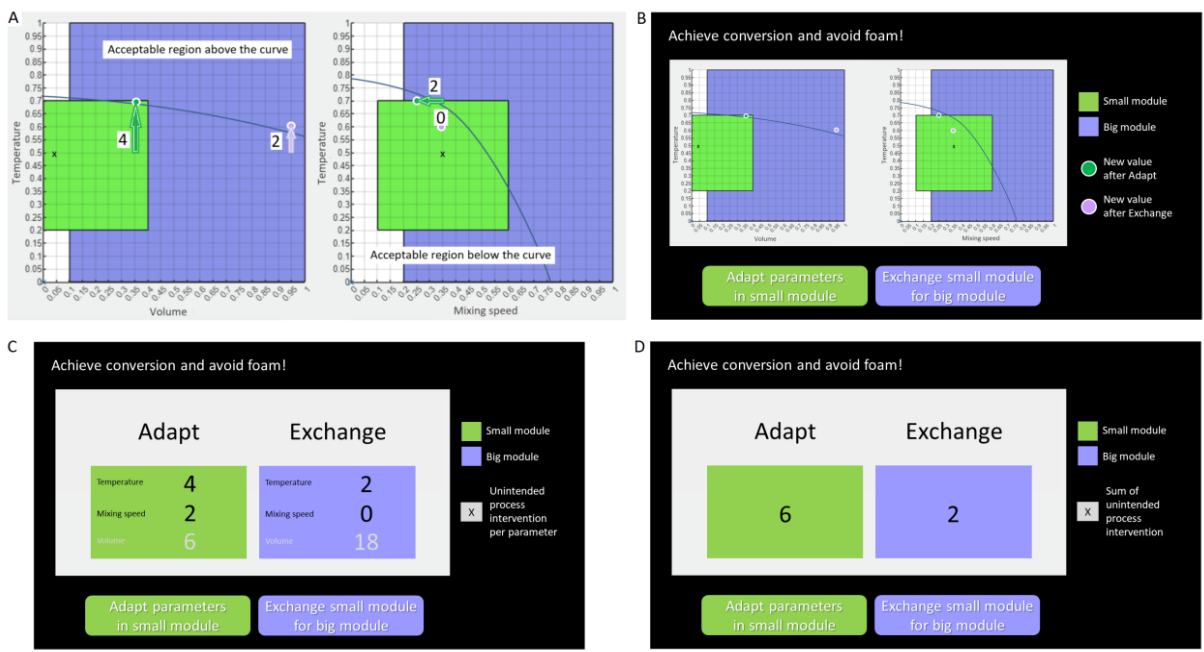

Graphs were shown in two adjacent pictures and visualised how the costs of the two solutions resulted from process parameter changes (see Figure 3B). Each picture reflected how one outcome depended on the interaction of two parameters: for conversion (left picture), the relation between temperature on the y-axis and volume on the x-axis reflected whether any given parameter value combination was able or unable to exceed the conversion threshold. This threshold was exceeded for all value combinations above the curve, while all value combinations below the curve were invalid as they did not achieve the required conversion. For foam (right picture), the relation between temperature on the y-axis and mixing speed on the x-axis reflected whether any given parameter value combination was able or unable to stay below the acceptable foam risk. All combinations below the curve met this foam requirement. The slopes and positions of the curves changed between trials. In all graphs, green and purple background boxes visualised the parameter ranges of the small and big module, respectively. These ranges were fixed throughout the experiment. The current parameter settings were marked by a black x in both pictures. Moreover, two solutions (i.e., new parameter values) were included in each graph, one for Adapt (green dot) and one for Exchange (purple dot). These dots were always positioned above the conversion curve and below the foam curve, thus reflecting valid solutions. They also reflected the minimal process intervention and thus the best possible outcome achievable with either solution. Their distance to the current parameter values (i.e., to the x) reflected the Adapt and Exchange costs. This distance was measured in the number of horizontal grid cells for temperature and vertical grid cells for mixing speed, which we will refer to as steps. One step corresponded to one grid cell (i.e., 0.5 units on the temperature or mixing speed scale). Volume was ignored and did not contribute to the step count, because increasing it had no costs.

Separate numbers represented the costs of Adapt and Exchange in terms of the process intervention needed for each individual parameter (see Figure 3C). Thus, these numbers also reflected the number of steps (i.e., grid cells to be travelled in the graphs). While temperature and mixing speed steps were presented in black font, volume steps were presented in grey font, as they were irrelevant and could be ignored. In stimuli with integrated numbers (see Figure 3D), the steps required for temperature and mixing speed were added up, and thus only a single number was shown for each solution. In both numeric formats, these steps or costs were represented in a green box for Adapt (left) and a purple box for Exchange (right).

Adapt always yielded higher costs than Exchange, but the stimuli differed in their absolute magnitude of Adapt costs (i.e., 3, 6 or 18 steps) and in the Adapt/Exchange ratio (i.e., 6:1, 6:2, or 6:3). Table 1 illustrates how the Exchange costs depended on Adapt costs and Adapt/Exchange ratio. For instance, in a trial with an Adapt cost of 18 steps and an Adapt/Exchange ratio of 6:2, Exchange required 6 steps.

**Table 1**

Exchange costs (steps) in Experiment 1, depending on Adapt costs and Adapt/Exchange ratio.

|  |  | Adapt/Exchange ratio | | |
|---|---|---|---|---|
|  |  | 6:1 | 6:2 | 6:3 |
| Adapt costs (steps) | 3 | 0.5 | 1 | 1.5 |
|  | 6 | 1 | 2 | 3 |
|  | 18 | 3 | 6 | 9 |

A total of 110 process graphs was generated, and each trial had its own stimulus. These stimuli differed in their assignment of total costs to temperature and mixing speed steps. For instance, an integrated cost of 6 steps could result from a temperature increase of 4 and a mixing speed decrease of 2 steps, or it could result from a change of both parameters by 3 steps. The stimuli also differed in the number of required volume steps, although volume was not included in the step calculations. The exact same step numbers that were represented in graphs were also used in the numeric presentation formats.

The *physical module exchange* was simulated using five Mega Bloks® (i.e., big, coloured plastic blocks) that were placed next to participants' laptops.

*2.2.3 Procedure*

*Instruction, practice, and knowledge test.* After being welcomed and signing the consent form, participants watched the instruction video. Afterwards, they completed five practice tasks on paper, in which they were shown graphs and had to generate a solution. These graphs only contained the current parameter values, but not the new, intended values. Participants had to choose a solution (Adapt of Exchange) and set the parameters volume, temperature, and mixing speed to their new values with a pen. In doing so, they had to make sure to get above the conversion curve and stay below the foam curve. Their solutions were subsequently checked by the experimenter. In case of errors, she provided feedback and used the erroneous examples to explain again how the parameter values had to be set. After the practice task, participants performed a written multiple-choice knowledge test to assess their understanding and memory of the instruction. While all participants were allowed to take part in the experiment, a criterion of 70 % correct answers was used to decide whether the data would be analysed. Taken together, the instruction, practice, and knowledge test took about 30-45 minutes.

*Experiment*. The experiment used a three-factorial within-subjects design: 3 (*presentation format: graphs, separate numbers, integrated numbers*) × 3 (*Adapt costs: 3, 6, 18 steps*) × 3 (*Adapt/Exchange ratio: 6:1, 6:2, 6:3*). During the experiment, participants performed nine practice trials (three for each presentation format) and 330 experimental trials. The latter consisted of 270 relevant trials: ten repetitions of each combination of the three factors (i.e., presentation format, Adapt costs, Adapt/Exchange ratio). Additionally, they included 60 filler trials: ten repetitions per presentation format of two irrelevant cost factor combinations (i.e., 3 Adapt steps with an Adapt/Exchange ratio of 6:6, and 18 Adapt steps with an Adapt/Exchange ratio of 6:0.33). These filler trials had been added to make the step distributions appear balanced to participants (by including 3 and 1 absolute Exchange steps in each Adapt cost condition). However, they were not included in the data analysis. Trial order was randomised across the experiment, individually for each participant. The experiment was split into six blocks, allowing participants to take breaks after each 55 trials.

Trials started with a choice screen that presented the graphs or numbers (see Figure 3B-D). Participants had to choose between Adapt and Exchange by clicking one of the two buttons. Clicking the 'Adapt' button led them to a neutral screen with only the background objects (i.e., grey field, buttons, and legend), but no graphs or numbers, and the next trial started after 300 ms. Clicking the 'Exchange' button transferred participants to an exchange screen that prompted them to perform the physical module exchange. To simulate this procedure, they had to re-stack a pile of five Mega Bloks® upside-down. The minimum time for the module exchange task was set to five seconds, and only after this interval the 'Finish' button became active. On average, it took participants 7.2 sec to complete the block stacking task. After clicking the 'Finish' button, the neutral screen was presented for 300 ms, and the next trial started. Altogether, the experiment took about 1.5 hours.

*2.2.4 Data analysis*

To test whether participants satisficed or compared solutions when using the three presentation formats, the mean percentage of Exchange choices was analysed, using a 3 (*presentation format: graphs, separate numbers, integrated numbers*) × 3 (*Adapt costs: 3, 6, 18 steps*) × 3 (*Adapt/Exchange ratio: 6:1, 6:2, 6:3*) repeated measures ANOVA. Only effects including the factor presentation format are reported, as only these effects are relevant regarding the purpose of the study. An alpha value of *p* = .05 was used to determine statistical significance, and all pairwise comparisons were performed with Bonferroni correction. All data are available in our repository at the Open Science Framework: https://osf.io/8c5mz/

## 2.3 Results

Overall, Exchange was chosen in 41.4 % of the trials, and Table 2 provides an overview of the means for each cell. When comparing the mean percentage of Exchange choices between the experimental conditions, all main effects and interactions were significant (see Figure 4). First, a main effect of presentation format, $F(2,48) = 7.375$, $p = .002$, $\eta_p^2 = .235$, indicated that Exchange was chosen less often with graphs than with separate and integrated numbers (31.4 vs. 45.6 and 47.3 %), both $p$s < .03, while the two numeric formats did not differ, $p > .9$. Second, a main effect of Adapt costs, $F(2,48) = 72.151$, $p < .001$, $\eta_p^2 = .750$, indicated that Exchange choices increased with Adapt costs (17.3, 34.0 and 72.9 % for 3, 6 and 18 Adapt steps, respectively), all $p$s < .002. Third, a main effect of Adapt/Exchange ratio, $F(2,48) = 33.206$, $p < .001$, $\eta_p^2 = .585$, indicated that Exchange was chosen most often with an Adapt/Exchange ratio of 6:1, followed by 6:2 and 6:3 (50.5, 41.6 and 32.1 %), all $p$s < .002.

The interaction of presentation format and Adapt costs, $F(4,96) = 4.701$, $p = .002$, $\eta_p^2 = .164$, indicated that the reduced Exchange rate with graphs was restricted to medium and high Adapt costs: With only 3 Adapt steps, the presentation formats did not differ, all $p$s > .4. With 6 steps, graphs led to fewer Exchange choices than integrated numbers, $p = .007$, but the difference to separate numbers missed the significance level, $p = .057$. With 18 steps, graphs led to fewer Exchange choices than both numeric formats, both $p$s < .02. Thus, when the process intervention required for Adapt was low, presentation format did not affect choice, but as Adapt costs increased, graphs seemed to discourage Exchange. No difference between the numeric formats was detected at any level of Adapt costs, all $p$s > .4.

Moreover, there was an interaction of presentation format and Adapt/Exchange ratio, $F(4,96) = 8.158$, $p < .001$, $\eta_p^2 = .257$. Exchange was chosen more often with the highest Adapt/Exchange ratio of 6:1 than with the lowest ratio of 6:3 for all presentation formats, all $p$s < .003. This ratio dependence was smaller with graphs than with separate numbers and integrated numbers (with differences of 9.9 vs. 17.9 and 27.6 %, respectively). Still, we found significant Adapt/Exchange ratio effects for graphs. This disconfirms the hypothesis that graphs generally induce satisficing, which should have rendered the Adapt/Exchange ratio irrelevant.

The triple interaction of presentation format, Adapt costs, and Adapt/Exchange ratio just reached significance, $F(8,192) = 1.997$, $p = .049$, $\eta_p^2 = .077$, reflecting that Adapt costs were critical in determining whether the presentation formats differed in their sensitivity to the Adapt/Exchange ratio. With low Adapt costs of 3 steps, no presentation format led to differences between the Adapt/Exchange ratios, all $p$s > .1. For all presentation formats, Adapt was chosen most of the time, regardless of how much better Exchange might be (which suggests that participants satisficed). Conversely, with high Adapt costs of 18 steps, the Adapt/Exchange ratio influenced choices regardless of presentation format: Exchange was always chosen more often at the highest than the lowest Adapt/Exchange ratio, all $p$s < .03 (which suggests that participants compared the solutions). Differences between the presentation formats in their sensitivity to the Adapt/Exchange ratio were only observed with medium Adapt costs of 6 steps. Here, the difference between the highest and lowest Adapt/Exchange ratio was present for both numeric formats, both $p$s < .003, but absent for graphs, $p = .198$. Any trend for Adapt/Exchange ratio effects with graphs at 6 Adapt steps that might seem apparent in Figure 4 was exclusively due to two participants with extreme values (see Figure 5A).

Despite these general effects, there was a high interindividual variation in participants' choices (see Figure 5). For instance, some participants switched between never choosing Exchange at 3 and 6 Adapt steps to always choosing it at 18 Adapt steps, especially with integrated numbers. In fact, several participants displayed ceiling effects at 18 Adapt steps, choosing Exchange in 100 % of the trials, regardless of Adapt/Exchange ratio. One participant even chose Exchange in all but one trial across the entire experiment, and others rarely chose it at all.

**Table 2**

Means and standard deviations (in parentheses) for the percentage of Exchange choices in Experiment 1, depending on presentation format, Adapt costs, and Adapt/Exchange ratio.

| Adapt costs | 3 steps | | | 6 steps | | | 18 steps | | |
|---|---|---|---|---|---|---|---|---|---|
| Adapt/Exchange ratio | 6:1 | 6:2 | 6:3 | 6:1 | 6:2 | 6:3 | 6:1 | 6:2 | 6:3 |
| Graphs | 16.0 (23.3) | 15.6 (26.3) | 13.6 (24.1) | 25.6 (32.8) | 22.8 (26.9) | 14.4 (25.8) | 65.6 (37.3) | 59.2 (36.7) | 49.6 (39.0) |
| Separate numbers | 25.6 (32.7) | 22.0 (30.7) | 17.6 (32.3) | 46.8 (35.3) | 36.0 (37.1) | 26.4 (35.6) | 89.6 (21.1) | 81.6 (30.4) | 64.4 (35.8) |
| Integrated numbers | 22.8 (34.6) | 12.4 (27.0) | 10.4 (21.7) | 68.4 (39.3) | 39.2 (40.2) | 26.4 (40.5) | 94.4 (13.9) | 85.6 (26.6) | 66.0 (37.5) |

**Figure 4**

Percentage of Exchange choices in Experiment 1, depending on presentation format, Adapt costs, and Adapt/Exchange ratio. Error bars represent standard errors of the mean.

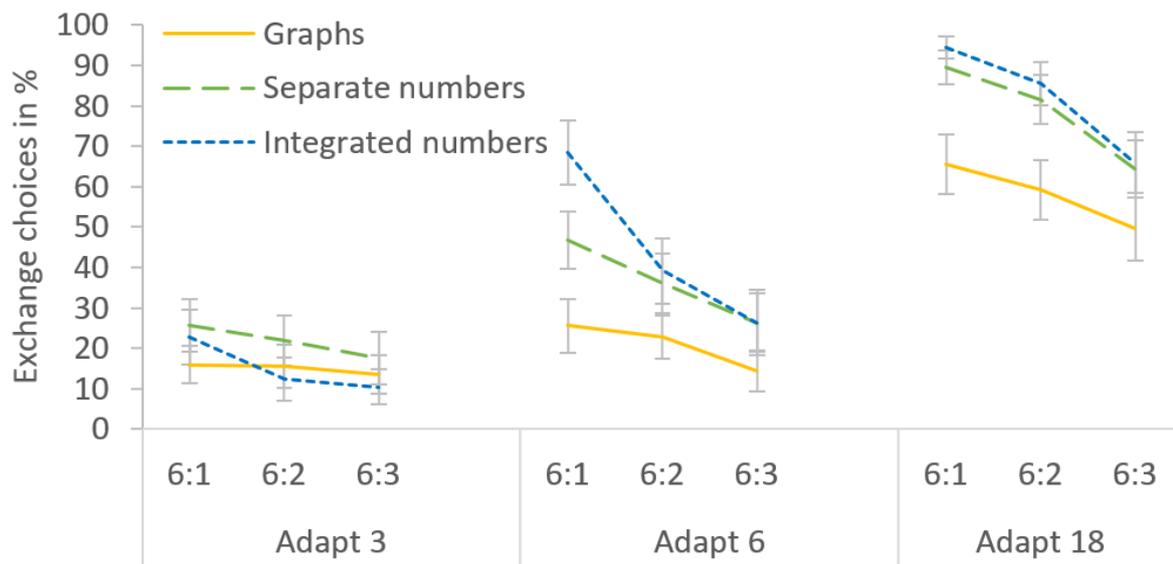

**Figure 5**

Individual percentages of Exchange choices in Experiment 1 for each presentation format, depending on Adapt costs and Adapt/Exchange ratio, with (A) graphs, (B) separate numbers, and (C) integrated numbers. Each line represents the data of one participant.

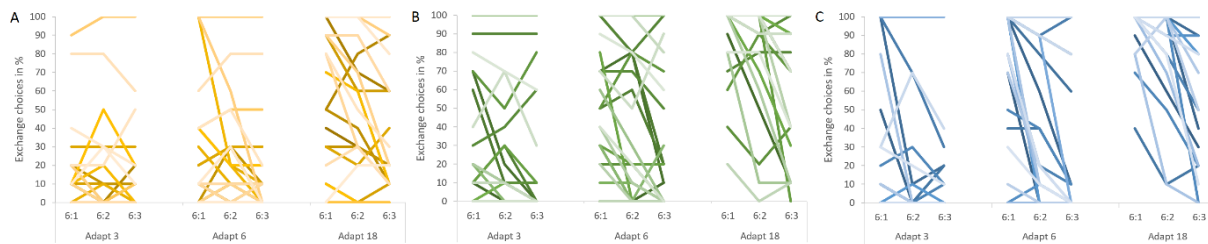

## 2.4 Discussion

Experiment 1 examined how strategies of making Adapt/Exchange decisions depend on the format in which the costs of both solutions are presented. It was hypothesised that graphs induce satisficing, because they pose higher demands on information search and integration, thereby making it harder to compare the solutions. Therefore, Exchange choices should be independent of the Adapt/Exchange ratio, which indicates satisficing. Conversely, a presentation of costs as numbers was hypothesised to facilitate solution comparison, leading to more Exchange choices with higher Adapt/Exchange ratios.

The picture that emerged from the results is more complex. The presentation formats indeed showed differential sensitivity to the Adapt/Exchange ratio, but only with medium Adapt costs. In this case, the effects supported our hypotheses. That is, with graphs, participants seemed to satisfice instead of comparing solutions: the frequency of Exchange choices did not depend on the Adapt/Exchange ratio. With both numeric formats, participants did compare the solutions: they chose Exchange more often when the Adapt/Exchange ratio was high. Thus, when the situation was ambiguous regarding the quality of the status quo (i.e., medium Adapt costs), presentation format mattered. A different picture emerged when Adapt costs got more extreme in either direction. With low Adapt costs, participants satisficed irrespective of presentation format: no presentation format revealed Adapt/Exchange ratio effects, and participants simply chose Adapt, regardless of whether Exchange led to better outcomes. With high Adapt costs, participants compared the solutions irrespective of presentation format: all presentation formats revealed Adapt/Exchange ratio effects. This is interesting, as it shows that participants were clearly capable of comparing the solutions with graphs. Thus, graphs did not make solution comparison impossible, but merely discouraged it when it was not necessary.

An unexpected finding was the main effect of presentation format: Participants chose Exchange less often with graphs than with both numeric formats. On the one hand, this might be a genuine consequence of the different presentation formats. On the other hand, it might have resulted from how the numbers were generated. They represented the steps needed to move above the conversion curve and below the foam curve in the process graphs, ranging from 0.5 to 18. However, the axes of the graphs did not represent steps, but standardised process units, ranging from 0 to 1. Accordingly, the numbers shown to participants were larger in the numeric formats.

The presentation formats were not necessarily inconsistent, because in fact, we do not know how participants represented the process intervention costs with graphs. Given the salient grid markings, it is likely that they thought of them as steps (i.e., in the same unit as the numeric formats). However, it is also possible that participants were aware of the difference in scale, and thus adjusted their behaviour accordingly. Previous research suggests this to be consequential. First, higher reward magnitudes in intertemporal choice make people more inclined to choose the larger/later reward (Green et al., 1997). Similarly, our higher absolute costs in the numeric formats might have discouraged participants from choosing the costly Adapt solution. Second, in multi-attribute decision making, expanded numeric scales (e.g., costs per year instead of per month) make people more inclined to choose the solution that performs better on the expanded attribute (Burson et al., 2009). Similarly, the disadvantage of Adapt might have seemed larger for the numeric formats. To control for such effects of inconsistent scales, in Experiment 2 the numbers were changed from steps to process units, so that the process intervention cost was measured on the same scale for all three presentation formats.

## 3. Experiment 2

### 3.1 Introduction

The aim of Experiment 2 was twofold. First, we intended to replicate the independence of choices from Adapt/Exchange ratio with graphs at medium Adapt costs. Second, we eliminated a confound of Experiment 1, where presentation formats had differed in the numeric scaling of costs. In Experiment 2, all presentation formats used the same scale to represent process interventions (i.e., 0-1). If the effects of presentation format in Experiment 1 had resulted from different scales, they should disappear. If they had actually resulted from graphs hampering the comparison of Adapt and Exchange, they should remain intact.

However, changing the scale of costs had another consequence: The numeric formats now had to use a varying number of decimal places. Accordingly, neither the magnitude of the single digits, nor the length of the string could be used by participants to evaluate the required process intervention (e.g., 0.1 is larger than 0.075, but the latter has higher digit values and is longer). Decimals are not processed automatically, and string length strongly influences their processing (Kallai & Tzelgov, 2014). The use of decimals can also affect decision making, and leads to less discounting in intertemporal choice (Fassbender et al., 2014). A higher difficulty in comparing the costs of the two solutions might decrease the dependence of choices on the Adapt/Exchange ratio. In Experiment 1, stronger Adapt/Exchange ratio effects were observed with numbers than graphs. If this was because numbers facilitated comparison, the difference between presentation formats should be reduced or absent in Experiment 2. Particularly, the Adapt/Exchange ratio effects for separate numbers should become more similar to graphs, because participants had to compare and additionally add decimal numbers.

Taken together, changing the numeric scale has two consequences: it lowers the cost magnitudes and increases the difficulty of comparing decimals. Both should reduce the differences between presentation formats. Particularly, the main effect of presentation format and the interaction of presentation format and Adapt/Exchange ratio should disappear, if they had only been a consequence of low-level factors in the scaling of numbers. Instead, if they reflected a genuine effect of presentation format, they should remain intact.

**3.2 Methods**

*3.2.1 Participants*

Twenty-eight members of the TUD Dresden University of Technology participant pool (ORSEE, Greiner, 2015) took part in the study in exchange for course credit or 7€ per hour. Three participants were excluded from the data analysis: one because she displayed a severe lack of understanding and could not even solve the practice task, and two because they did not pass the knowledge test. Thus, the final sample consisted of 25 participants (15 female, 10 male) with an age range of 19 to 47 years (*M* = 27.3, *SD* = 7.2). Participants provided written informed consent, and all procedures followed the principles of the Declaration of Helsinki.

*3.2.2 Apparatus and stimuli*

The stimulus material was identical to that of Experiment 1, with the following exception: The numeric formats now represented units of process intervention instead of steps, and thus matched the scale of the graph axes, ranging from 0 to 1. To this end, all numbers from Experiment 1 were divided by 20. Note that this resulted in decimal numbers with varying string length (see Table 3).

**Table 3**

Absolute costs of Exchange in Experiment 2, depending on Adapt costs and Adapt/Exchange ratio.

|  |  | Adapt/Exchange ratio | | |
|---|---|---|---|---|
|  |  | 6:1 | 6:2 | 6:3 |
| Adapt costs (units) | 0.15 | 0.025 | 0.05 | 0.075 |
|  | 0.3 | 0.05 | 0.1 | 0.15 |
|  | 0.9 | 0.15 | 0.3 | 0.45 |

*3.2.3 Procedure*

The procedure was identical to that of Experiment 1.

*3.2.4 Data analysis*

The methods of analysing the data were identical to Experiment 1, and only the levels of the Adapt cost factor in the repeated measures ANOVA were different: 3 (*presentation format: graphs, separate numbers, integrated numbers*) × 3 (*Adapt costs: 0.15, 0.3, 0.9 units*) × 3 (*Adapt/Exchange ratio: 6:1, 6:2, 6:3*). All data are made available via the Open Science Framework: https://osf.io/8c5mz/

**3.3 Results**

Overall, Exchange was chosen in 41.1 % of the trials, and Table 4 provides an overview of the means for each cell. When comparing Exchange choices between the experimental conditions (see Figure 6), a first striking result was that the main effect of presentation format was completely absent, $F < 1$. Exchange choices did not differ between graphs, separate numbers, and integrated numbers (37.5 vs. 41.8 and 43.9 %), all $ps > .9$. Second, a main effect of Adapt costs, $F(2,48) = 60.541$, $p < .001$, $\eta_p^2 = .716$, indicated that Exchange was chosen more often when Adapt was more costly (23.0, 32.1 and 68.0 % for 0.15, 0.3 and 0.9 Adapt units, respectively), all $ps < .001$. Third, a main effect of Adapt/Exchange ratio, $F(2,48) = 16.362$, $p < .001$, $\eta_p^2 = .716$, indicated that Exchange was chosen most often with an Adapt/Exchange ratio of 6:1, followed by 6:2 and 6:3 (50.0, 40.1 and 33.2 %), all $ps < .007$.

The interaction of presentation format and Adapt costs was not significant, $F(4,96) = 1.628$, $p = .173$, $\eta_p^2 = .064$, indicating that presentation formats did not differ for any level of Adapt costs, all $ps > .3$. However, there was a significant interaction of presentation format and Adapt/Exchange ratio, $F(4,96) = 3.023$, $p = .021$, $\eta_p^2 = .112$. Although Exchange was chosen more often with an Adapt/Exchange ratio of 6:1 than with 6:3 for all presentation formats, all $ps < .02$, this difference between the highest and lowest Adapt/Exchange ratio was smaller with graphs than with separate numbers or integrated numbers (9.5 vs. 17.9 and 22.3 %). However, again the data do not support the hypothesis that Adapt/Exchange ratio is generally irrelevant with graphs.

The triple interaction of presentation format, Adapt costs and Adapt/Exchange ratio did not reach significance, $F(8,192) = 1.722$, $p = .096$, $\eta_p^2 = .067$. However, pairwise comparisons indicated that Adapt costs determined whether the presentation formats differed in their sensitivity to the Adapt/Exchange ratio. With low Adapt costs, the difference between the highest and lowest Adapt/Exchange ratio was significant for graphs and separate numbers, both $ps < .05$, but not for integrated numbers, $p = .089$. With medium Adapt costs, this Adapt/Exchange ratio effect was absent for graphs, $p = .787$, but present for both numeric formats, both $ps < .02$. With the highest Adapt costs, the Adapt/Exchange ratio effect missed the significance level for graphs, $p = .068$, but was highly significant for both numeric formats, both $ps < .005$. Thus, similar to Experiment 1, the most pronounced influence of presentation format on whether participants compared the solutions was observed with medium Adapt costs: when using graphs, it did not matter whether Exchange was much better than Adapt, while when using the numeric formats, it did. Just like in Experiment 1, there were considerable interindividual differences in the percentage of Exchange choices (see Figure 7).

**Table 4**

Means and standard deviations (in parentheses) for the percentage of Exchange choices in Experiment 2, depending on presentation format, Adapt costs, and Adapt/Exchange ratio.

| Adapt costs | 0.15 units | | | 0.3 units | | | 0.9 units | | |
|---|---|---|---|---|---|---|---|---|---|
| Adapt/Exchange ratio | 6:1 | 6:2 | 6:3 | 6:1 | 6:2 | 6:3 | 6:1 | 6:2 | 6:3 |
| Graphs | 25.2 (26.8) | 14.8 (17.6) | 12.4 (20.3) | 32.0 (32.9) | 33.2 (32.2) | 26.0 (31.9) | 68.8 (37.2) | 65.6 (37.1) | 59.2 (35.6) |
| Separate numbers | 32.8 (33.5) | 26.8 (33.4) | 21.6 (34.7) | 43.2 (41.0) | 31.2 (34.9) | 27.6 (35.7) | 77.6 (29.5) | 64.4 (33.9) | 50.8 (36.4) |
| Integrated numbers | 33.6 (41.7) | 20.0 (33.4) | 20.0 (33.5) | 44.8 (42.8) | 29.2 (38.3) | 22.0 (34.4) | 90.0 (23.5) | 76.0 (38.0) | 59.6 (43.5) |

**Figure 6**

Percentage of Exchange choices in Experiment 2, depending on presentation format, Adapt costs, and Adapt/Exchange ratio. Error bars represent standard errors of the mean.

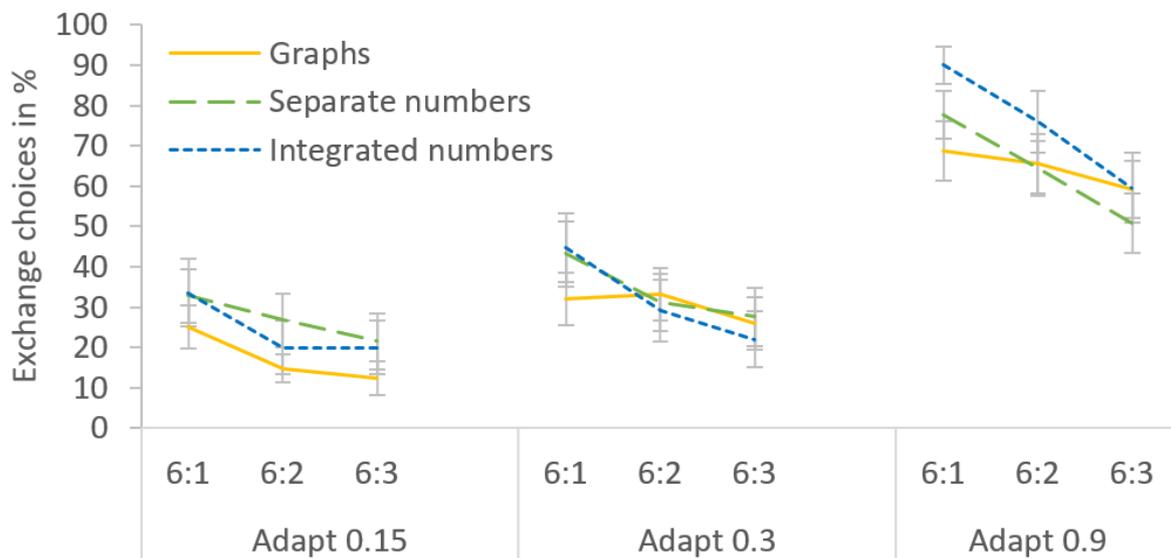

**Figure 7**

Individual percentages of Exchange choices in Experiment 2 for each presentation format, depending on Adapt costs, and Adapt/Exchange ratio, with (A) graphs, (B) separate numbers, and (C) integrated numbers. Each line represents the data of one participant.

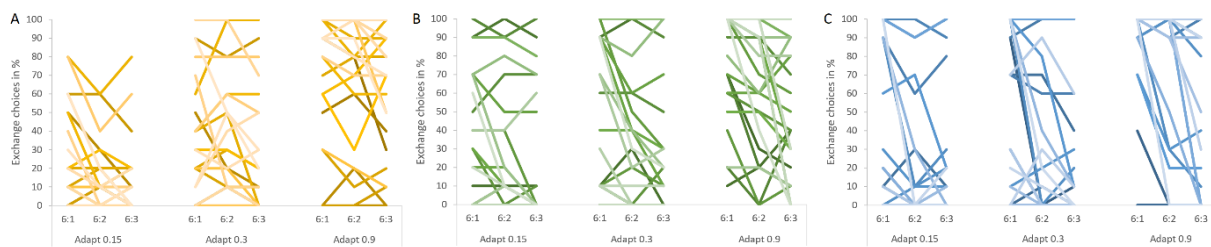

### 3.4 Discussion

Experiment 2 aimed to replicate the main findings of Experiment 1, while eliminating a potential confound in the scaling of numbers. Therefore, numbers were presented on the same scale as graph units. This indeed made participants choose Exchange similarly often with all presentation formats. However, this was not only due to a decrease of Exchange choices with the numeric formats, but a concurrent increase with graphs. A reason for this unexpected result might be that participants in Experiment 2 matched their Exchange choices between presentation formats more closely, because the identical scale now made it more obvious that all formats represented the same thing.

The most important finding of Experiment 2 was that presentation format still interacted with Adapt/Exchange ratio, suggesting that participants were less likely to compare the solutions with graphs than numbers (as indicated by weaker ratio effects for graphs). For the numeric formats, ratio effects remained intact, despite the altered number scale (including decimal places and different string lengths), which should have made solution comparison more difficult. Similar to Experiment 1, the difference between presentation formats was most pronounced with medium Adapt costs. In this condition, ratio effects were completely absent with graphs, but substantial with the numeric formats. The finding that in ambiguous situations, participants compare solutions with numbers but not with graphs, thus seems to be robust.

A curious finding from Experiment 2 was that in contrast to Experiment 1, graphs produced Adapt/Exchange ratio effects when only a small process intervention was necessary (i.e., with low Adapt costs). Currently, it is unclear how to interpret this result. However, given that for graphs the stimulus material did not differ from Experiment 1 in any way, it might well be a chance finding and should be replicated before drawing conclusions.

# 4. General discussion

In many situations, problem solvers have to choose between local and global changes: Should they merely adapt a currently used solution, or exchange it for a completely new one? Previous research suggested that people do not thoroughly compare these solutions, but merely check whether the status quo is good enough (Müller & Pohl, 2023; Müller & Urbas, 2020). But how does this selection of decision strategies depend on the constraints of the decision context, or features of the task? The present study investigate the role of structural features (i.e., solution costs) and surface features (i.e., presentation format) of Adapt/Exchange decisions. In two experiments using a modular plant scenario, participants either had to base their decisions on a graph visualisation of process relations, on numeric information about the costs for each process parameter, or on an integrated presentation of the total costs of each solution. Both experiments revealed that presentation format does indeed affect whether people satisfice or base their choices on differences between the costs of the two solutions. However, this format dependence was most pronounced when the situation was ambiguous regarding the costs of the status quo. In contrast, people were generally less likely to perform a thorough comparison of solutions when the status quo only had minimal costs, and more likely when the status quo was very costly. These findings emphasise that decision strategies are flexible and depend on an interplay between structural and surface features of the decision context.

## 4.1 How do Adapt/Exchange decisions depend on presentation format?

In the present study, the effects of presentation format were substantial. How solution costs were presented had an influence on decision strategies, despite being varied in random order, so that graphs were interspersed with the numeric formats. This experimental setup puts presentation format effects to a critical test, as it emphasises that in fact all three formats reflected the same choice situation. Indeed, several participants remarked during debriefing that they had found it quite striking to observe themselves making different choices with different presentation formats, despite knowing that the situation was the same. Still, the robustness of presentation format effects does not mean that a particular presentation forced participants into using a particular strategy. With graphs, participants also compared solutions (as indicated by Adapt/Exchange ratio effects), namely when Adapt costs were high. With the numeric formats, they also satisficed (as indicated by an absence of Adapt/Exchange ratio effects), namely when Adapt costs were low. Thus, surface features like presentation format only seem to encourage particular decision strategies when the circumstances are favourable. This is in line with the notion that the structure of the environment determines the selection and implementation of decision strategies (Todd et al., 2012).

It is remarkable that graphs, which arguably provided the most information, promoted the most "quick and dirty" decision strategy (i.e., satisficing instead of comparing solutions). This is interesting, because all information provided by the numeric formats was also included in the graphs, along with ample additional information (e.g., conversion and foam thresholds, current and new parameter values, distance from the module boundaries). At least three explanations are conceivable. First, satisficing with graphs might have resulted from *information overload* (Eppler & Mengis, 2004), as graphs might have given participants a hard time extracting the relevant information. This also is in line with the finding that a use of heuristics in multi-attribute decision making increases with task complexity (Timmermans, 1993) and with the need to search and integrate information (Glöckner & Betsch, 2008; Söllner et al., 2013). Second, satisficing might have resulted from the *specific information contents* of graphs. For instance, they made the module boundaries visible, perhaps emphasising that all solutions could be realised within these boundaries. In principle, it had also been made explicit in the instructions that all solutions were valid, but perhaps it became more salient with graphs. This is in line with design guidelines suggesting that clearly visible boundaries (i.e., constraints as containers) facilitate situation evaluation (Hansen, 1995). Third, satisficing might not be a "quick and dirty" strategy after all, but make perfect sense once the scenario exceeds a certain degree of complexity. Graphs presumably increased the *saliency of the complex industrial domain*, compared to numbers that could represent anything in principle (cf. Peters et al., 2009). Thus, participants might have been reminded of the complex, non-explicit domain constraints beyond the mere Adapt/Exchange ratio. Accordingly, this ratio might no longer have been sufficient to guide their choices.

All four explanations rest on the assumption that the additional information provided in graphs drew participants' attention away from the explicit costs. However, they make different assumptions about how the information was processed and how this induced satisficing. The first explanation assumes that the additional information exceeded participants' cognitive limitations. Satisficing is thus ascribed to excessive demands. The second explanation assumes that the specific information contents made participants evaluate all solutions as equally feasible. Satisficing is thus ascribed to indifference. The third explanation assumes a more abstract effect of the information, reminding participants that the explicit costs do not sufficiently constrain choices. Satisficing is thus ascribed to an ecologically rational consideration. Based on the present data, we cannot distinguish between the explanations. Moreover, they are not mutually exclusive, and combinations are conceivable. For instance, combining the first and third explanation would hold that the information load of the graphic presentation shifted participants' *weighting of explicit versus non-explicit costs*. This would resonate with findings that when quantitative information is hard to evaluate, people assign more weight to additional, non-quantitative information (Peters et al., 2009). Future studies should disentangle the mechanisms of why more information can lead to reductionist strategies in Adapt/Exchange decisions.

While the results obtained with both numeric formats differed from graphs, they were strikingly similar to each other. Although separate numbers required participants to integrate different information elements to compute the total solution costs, this did not manifest in different decisions than presenting the costs in an integrated manner right away. These findings contrast with intertemporal choice studies, where a segregation of numbers altered participants' choice behaviour (e.g., Grace & McLean, 2005; Magen et al., 2008; Radu et al., 2011). This inconsistency could stem from the fact that segregation made the options more distinguishable in intertemporal choice, whereas it hampered their comparison in the present study. However, the findings also contrast with multi-attribute decision making studies, where higher integration demands fostered heuristic strategies (Söllner et al., 2013). The latter inconsistency could stem from the fact that in the present study, the integration demands of separate numbers were quite low, only requiring an addition of two values per solution. Accordingly, our presentation format with high integration demands (i.e., separate numbers) still posed lower integration demands than the formats with low integration demands in previous studies.

**4.2 Conceptual questions**

A number of conceptual questions should be considered when evaluating the present findings. First, our *conceptualisation of satisficing* is somewhat different from other conceptualisations used in the literature. Typically, accounts of satisficing focus on decisions between many options (e.g., Gigerenzer et al., 2012; Payne et al., 1988; Simon, 1956; Todd & Miller, 1999), a search of large information spaces (e.g., Agosto, 2002; Prabha et al., 2007), or non-routine problem solving under severe time constraints (Klein, 2008). In these contexts, people have to eliminate options and truncate their search at some point. Conversely, in the present study, only two options (i.e., solutions) were available, and satisficing was conceptualised as choosing mere adaptations of the solution when it was good enough. Thus, not all of the present findings might generalise to other instances of satisficing, and vice versa. To bridge these gaps, it would be interesting to investigate how presentation formats influence at which point people stop searching for solutions in Adapt/Exchange scenarios. This is a relevant question in modular plants, because digital transformation allows making previous solutions available in large databases that can be searched when a similar problem reoccurs (Müller et al., 2021). In such case-based reasoning, the selection of suitable solutions is a major challenge (Lopez de Mantaras et al., 2005). Thus, it would be important to know whether more information-laden presentation formats (like the present graphs) discourage people from comparing past solutions to their current one, and prompt them to truncate their search prematurely. However, it would be equally problematic if simplistic presentation formats made people restrict their focus to incomplete explicit information and neglect the broader production context.

A second conceptual question is whether people really do not compare solutions, or simply *choose not to act upon this comparison*. A person might well be aware of the fact that Exchange is clearly superior in terms of its process intervention costs, but then consider this irrelevant, given the negligible costs of Adapt, or the large non-explicit costs of Exchange. A central advantage of humans over computers is that they can choose to ignore the data and base their judgments on information beyond what is made explicit (de Winter & Dodou, 2014). Contradicting the assumption that participants did make the comparison but then did not enact it, a recent study traced participants' information acquisition processes in Adapt/Exchange decisions (Müller & Pohl, 2023). It was found that the Exchange solution was less likely to be checked when Adapt was more feasible. Moreover, aside from a few exceptions, the majority of participants did not check Exchange when it was clear that Adapt could be applied without any problems. Future studies should continue to specify the cognitive mechanisms underlying Adapt/Exchange decisions, in order to find out at which stage of information processing alternative solutions cease to be considered in different contexts.

A related conceptual question is how the two *solutions are mentally represented*. For instance, we often referred to Adapt as "staying with the status quo", but at the same time it could be conceptualised as "avoiding effort" or "minimising the risk of harming the plant". These alternative perspectives differ in their approach versus avoidance focus, and should be subject to different biases such as loss aversion (Tversky & Kahneman, 1991), probability discounting (Green & Myerson, 2004), or effort discounting (Kool et al., 2010). As different types of discounting involve different decision processes (Białaszek et al., 2019), it would be interesting to know how different types of framing affect

Adapt/Exchange decisions. The intertemporal choice literature suggests that such framing is highly effective in changing decisions between complementary options (for an overview see Lempert & Phelps, 2016). Moreover, probability discounting is less sensitive to presentation format effects than delay discounting (Asgarova et al., 2017; Shead & Hodgins, 2009). Accordingly, a more explicit framing of Adapt and Exchange costs as risks might alter the impacts of presentation formats.

### 4.3 Limitations and outlook

Two opposite types of limitations should be considered. The first one concerns the internal validity of our experimental setting. Perhaps most importantly, only a fraction of the information about the two solutions was *explicit and quantified*. For instance, participants learned that temperature was particularly important, but were not instructed how to weight the parameters. Similarly, a large part of the Exchange costs was non-explicit. The physical effort associated with this solution could only be experienced by performing the manual Exchange procedure. Other costs were not perceivable at all, but had merely been explained to participants during the instruction. In consequence, it was unclear how costly Exchange actually was, and different participants might have had different ideas about this. On the one hand, such multidimensionality and intransparency of costs and benefits certainly reduces experimental control. The interindividual variability in our data clearly affirms this concern. On the other hand, these very features are characteristic of complex systems like modular plants. To assess how they affect people's decisions, it is inevitable to sacrifice experimental control to some extent. That said, future studies should systematically investigate the interplay between explicit and non-explicit costs. By varying the type and severity of non-explicit costs, one might assess how this changes people's information integration and strategy selection.

The second type of limitations concerns the present study's external validity. At the outset of the article, we highlighted the appropriateness of satisficing in complex environments such as modular plants. However, the complexity of our *simplified modular plant scenario* was quite low. This scenario neither included qualitatively different risks (e.g., harming the plant, harming the product, not being finished in time), nor did it confront participants with dynamic changes (e.g., processes getting progressively destabilised, delayed effects of Adapt and Exchange). Instead, participants had to make several hundreds of simplistic decisions. This might have led to routine and fatigue effects. More importantly, it certainly is not compatible with a careful deliberation of solutions based on an integration of qualitatively different costs and benefits. Thus, it is questionable which aspects of our findings will generalise to real-world Adapt/Exchange decisions.

Another consequence of the limited complexity is that our *graph visualisations did not bring any benefits*, but only made the comparison of solutions more difficult. Conversely, in real modular plants, the information provided by graphs can be essential. For instance, people should certainly know how close they are to the boundaries of a module's operating ranges, or by how much a solution exceeds the acceptability thresholds. If the information contained in the graphs had been more consequential, the effects of graphs to induce satisficing might have been mitigated or even reversed. For instance, graphs could favour one or the other solution by directing attention to attributes that differentiate between the solutions, similar to presentation format effects in intertemporal choice (e.g., LeBoeuf, 2006; Naudé et al., 2018; Radu et al., 2011). In this way, presentation formats might exert a wide variety of different influences on Adapt/Exchange decisions, depending on the specific information they provide.

As a third consequence of the simplistic setup, *Adapt and Exchange appeared quite similar* from a participant perspective. The qualitatively different nature of these two solutions did not become particularly transparent. Participants still seemed to take the difference between the solutions quite seriously, as indicated by a high Exchange rate of 41 % – even though this meant that participants also had to perform the physical block stacking task 137 times on average. However, this obviously does not mean that they mentally represented Adapt and Exchange as qualitatively different solutions.

Finally, it is without question that the available information shapes decision strategies in the real world. However, it is not always clear how. A key to understanding these dependencies is to analyse the cognitive requirements of similar decisions in different domains (Schmidt & Müller, 2023). For complex industrial systems like modular plants, this may require a close cooperation between psychologists and engineers. In this way, psychological studies can be derived from the actual problem structures that exist out there in world. This will strengthen our understanding of how people achieve a situation-specific balance between stability and flexibility in ecologically valid decision contexts.

# 5. Acknowledgments

Parts of this work were funded by grants of the German Federal Ministry for Education and Research (BMBF, grant number 02K16C070) and of the German Research Foundation (DFG, PA 1232/12-3). I want to thank Leon Urbas for his invaluable support in developing the modular plant scenario, and for many inspiring discussions about Adapt/Exchange decisions.